\RequirePackage{fix-cm}
\documentclass[smallextended, referee]{svjour3}       
\smartqed  
\usepackage{graphicx}
\usepackage{url}
\usepackage{hyperref}
\usepackage{lineno}
\usepackage{soul}
\usepackage{color}
\newcommand{\gst}{Ge$_2$Sb$_2$Te$_5$~}
\newcommand{\gstns}{Ge$_2$Sb$_2$Te$_5$}
\newcommand{\sn}{Si$_3$N$_4$~}
\newcommand{\snns}{Si$_3$N$_4$}
\newcommand{\oc}{$^{\circ}\mathrm{C}$}
\begin{document}
\title{Inter-diffusion of Plasmonic Metals and Phase Change Materials}


\author{Li Lu \and Weiling Dong \and  Jitendra K. Behera \and Li Tian Chew \and Robert E. Simpson}


\institute{Li Lu \at
              Singapore University of Technology and Design, 8 Somapah Road, Singapore, 487372 \\
              \email{li\_lu@mymail.sutd.edu.sg} 
           \and
           Weiling Dong \at
              Singapore University of Technology and Design, 8 Somapah Road, Singapore, 487372 \\
              \email{weiling\_dong@mymail.sutd.edu.sg} 
	     \and
           Jitendra K. Behera \at
              Singapore University of Technology and Design, 8 Somapah Road, Singapore, 487372 \\
              \email{jitendra\_behera@mymail.sutd.edu.sg} 
           \and
	    Li Tian Chew \at
              Singapore University of Technology and Design, 8 Somapah Road, Singapore, 487372 \\
              \email{litian\_chew@sutd.edu.sg}
           \and
           Robert E. Simpson \at
              Singapore University of Technology and Design, 8 Somapah Road, Singapore, 487372 \\
              Tel.: +65 64994519 \\
              \email{robert\_simpson@sutd.edu.sg}
}

\date{Received: date / Accepted: date}

\maketitle

\begin{abstract}
This work investigates the diffusion of metal atoms into phase change chalcogenides, which is problematic because it can destroy resonances in photonic devices.
Interfaces between \gst and metal layers were studied using X-ray reflectivity (XRR) and reflectometry of metal--\gst layered stacks.
The diffusion of metal atoms influences the crystallisation temperature and optical properties of phase change materials.
When Au, Ag, Al, W  structures are directly deposited on \gstns, inter-diffusion occurs. 
Indeed, Au reacts with \gst{} to form a AuTe$_2$ layer at the interface.
Diffusion barrier layers, such as \sn or stable  plasmonic materials, such as TiN, can prevent the interfacial damage.
This work shows that the interfacial diffusion must be considered when designing phase change material tuned photonic devices, and that TiN is the most suitable plasmonic material to interface directly with \gstns.
\end{abstract}

\section{Introduction}
\label{intro}
Manipulating light on the nanoscale using tuneable metamaterials has a range of important applications from detecting minute concentrations of bio-markers to high resolution printing\cite{sreekanth2018ami,Kumar12NN}.
Once these metamaterials have been fabricated their optical response is usually fixed, and this lack of adaptability limits the potential of the devices.
Integrating chalcogenide phase change materials (PCM) into photonic devices offer a potential route to reprogrammable photonic devices\cite{Cao13JOSAB_1,Dong18AOM}  

Phase change chalcogenides have amorphous and crystalline phases that are stable at room temperature.
There is a   a large optical and electrical contrast between these structural  phases. 
Switching between the states is very fast; indeed  femtosecond, picosecond, and nanosecond  electrical or laser pulses have been used to switch phase change materials\cite{Waldecker2015nmat,Loke12Sci,Behera17OME}.
The large optical property contrast has been commercialised in rewritable optical data storage discs \cite{Wuttig2007} whilst the large electrical property contrast is now being applied to non-volatile memory \cite{burr2010phase}.
The next chapter of phase change materials (PCM) research will involve developing materials and devices that exploit both the optical and electrical property changes together.
Indeed, tuneable polarisation-independent perfect absorbers at visible \cite{Cao2014} and mid-infrared frequencies \cite{Cao2013,tittl2015switchable}, reflective-displays \cite{Hosseini2014}, on-chip photonic memory\cite{Rude2015,Rios15NP}, and reconfigurable optical circuits \cite{rude2013optical,Stegmaier17AOM} have already been demonstrated.
However, most research groups do not consider interfacial reactions and diffusion between the phase change materials and metal layers\cite{Raoux2009}. 
We find that this is, actually, a big problem and must be considered when designing phase change material tuned plasmonic structures.
Diffusion can influence  the crystallisation kinetics, and optical constants of the phase change materials\cite{Piccione2013,Pandian2006,Kozyukhin2011,Dong16TSF}.
For this reason phase change optical data storage digital versatile discs (DVD) usually protect  Sb-Te based chalcogenide layers with a thin layer of ZnS-SiO$_2$\cite{Ohshima1996,cheng2009influence}.
Despite the diffusion problem, many recent phase change material photonics publications neglect the diffusion barrier.
This leads to uncertainty in the route cause of the optical strcture's switching response.  

From a plasmonic design perspective, placing the chalcogenide directly in contact with the metal is ideal.
This is because plasmonic effects decay exponentially with distance, and therefore the change to the optical response depends on the distance between the metal layer and the PCM.
In particular, tuneable hyperbolic metamaterials, which have hyperbolic dispersion, can be made by combining metals and chalcogenides as multilayer stacked structures\cite{lu2018JO}.
The layer thickness is typically $\sim$10 nm and the large number of interfaces means that inter-diffusion over nanometer distances can be a real problem.
Thus, it is important to identify plasmonic materials that can directly interface with \gstns, and other important phase change chalcogenides\cite{Dong18}, without inter-diffusion between the metal and the PCM layers occuring.

\section{Materials and methods}
\label{method}
\subsection{Fabrication of the metal/\gst layered stacks}
\label{sec:1}
The different metal layers, Ag, Al, Au, W, and TiN, were deposited on the silicon substrate  by Radio Frequency (RF) sputtering.
The chamber base pressure was 5.7 $\times$ 10$^{-5}$ Pa.
The deposition pressure was 0.5 Pa.
Ag, Al, Au, and W were deposited in an Ar atmosphere.
The deposition rate was 0.87 $\mathring{A}$/s (50 W, 230 s) for Ag,  0.8 $\mathring{A}$/s (100 W, 247 s) for Al, 1.8 $\mathring{A}$/s (50 W, 110 s) for Au, and 1.3 $\mathring{A}$/s (80 W, 154 s) for W.
The TiN layer was made by reactively sputtering Ti in an Ar : N$_2$ atmosphere of 5 : 5.
The deposition rate was 0.17 $\mathring{A}$/s (100 W, 1200 s).
The thickness of the metal layer was 20 $\pm$ 2 nm.
Then a 20~nm thick \gst{} layer was deposited on the metal layer with a deposition rate of 0.86 $\mathring{A}$/s (30 W, 233 s).
We also investigated the effect of \sn diffusion barriers between the \gst{} and metal layers. 
The \sn layers were RF sputtered from a Si target in a Ar : N$_2$ = 8:2 atmosphere at a pressure of 0.5 Pa.
The deposition rate was 0.068 $\mathring{A}$/s (45 W, 736 s).
In all cases, the sputtering target diameter was 50 mm.

\subsection{Interface  characterisation}
\label{sec:2}
X-ray reflectivity (XRR) was chosen to characterise the interfaces of the multilayer samples.
Since the XRR X-ray beam is spread over a cm$^2$-scale area, it provides a more representative description of the interface roughness and layer inter-diffussion than locallised nanoscale characterisation techniques.
The interference of the X-rays  reflected from the interfaces results in oscillations as a function of the X-ray beam incident angle. 
A model based on Parratt's equations can be fitted to these oscillations\cite{als2011elements}.
Typically the XRR intensity changes by four or five orders of magnitude over the scanned angular range.
XRR is highly sensitive to diffusion and roughness.
In particular, roughness decreases the reflected intensity because it causes diffuse scattering \cite{gibaud2000x}.

\subsection{FDTD simulations}
\label{sec:3}
The reflectance spectrum for the structure Au/\gstns/Au structure without and with a TiN diffusion barrier was modelled using the finite-difference time-domain (FDTD) method to solve Maxwell's equations \cite{fdtdlumerical}.
In the model periodic boundary conditions were used on the lateral dimensions with a size of 1000~nm by 1000~nm, perfectly matched layer boundary conditions were used in the vertical direction.
A plane wave source illuminated the layered structure at normal incidence.
The optical constants of gold described by Palik were used \cite{palik1997handbook}, whilst our ellipsometry measurements for the optical constants of our \gst films were used for the PCM layer\cite{chew2017chalcogenide}. 
Our \gst{} optical constants are available to download\cite{actaGSTnk}.

\section{Results and discussion}
\label{results}
We studied inter-diffusion at the interface between \gst and five different metals: Ag, Al, Au, W, and TiN.
Ag, Al, and Au are commonly used metals  in visible plasmonics due to their real part of the permittivity being less than zero at visible and near-infrared frequencies.
In addition we also studied W and TiN, which are often used in phase change random-access memory (PCRAM) devices as electrodes and contact directly with \gst \cite{burr2010phase}.
Therefore, we expect diffusion of W and TiN to be less problematic than the noble metals.
TiN is particularly attractive because its real composonent of the refractive index is less than zero for visible wavelengths of light. 
Indeed, its dielectric function is similar to that of Au \cite{naik2012titanium} and our as-deposited TiN films consequently looked golden.
Other metals such as Mo have also been used to contact directly with phase change material in  RF-devices \cite{wang2014low}.
In this paper, however, we focus on visible plasmonic metals, such as Au, Ag, TiN, and Al because our principle aim  is to alert those designing visible and near-infrared active plasmonics devices that the interfaces between many plasmonic metals and phase change materials, such as \gstns, are unstable. 

\subsection{Inter-diffusion at metal--\gst{} interfaces}
\label{sec:4}
\begin{figure}[htbp] 
   \centering
   \includegraphics[width=0.8\columnwidth]{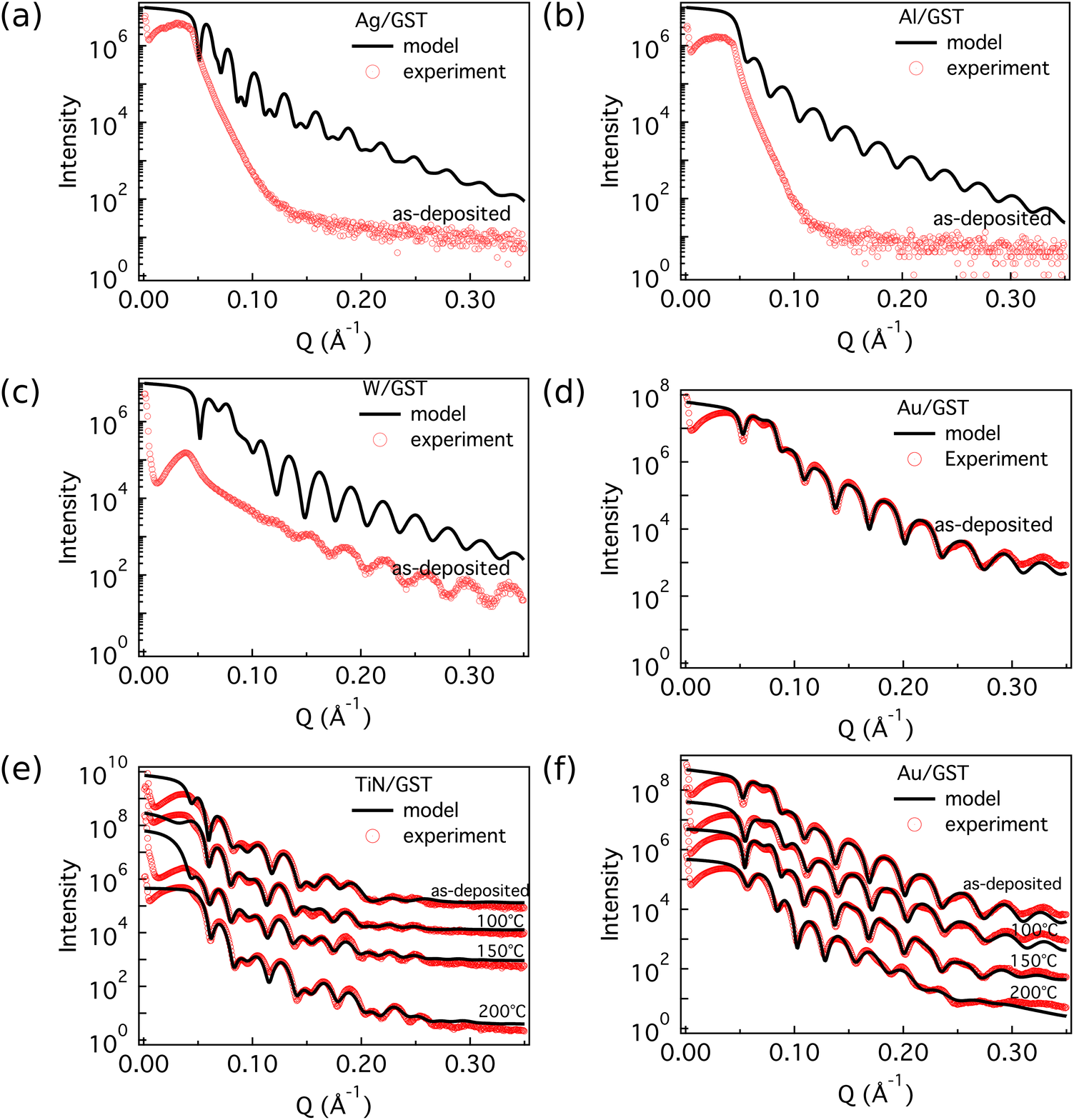} 
   \caption{XRR measurement and modelled curves for (a) Ag/\gstns, (b) Al/\gstns, (c) W/\gstns, and (d) Au/\gst in the as-deposited state; (e) TiN/\gstns, and (f) Au/\gst in the as-deposited state and after annealing at 100 \oc, 150 \oc~ and 200 \oc.}
   \label{fig_XRR}
\end{figure}

We chose the XRR method to study the buried thin film interfaces because it gives a global measurement of the interface and it is not limited to a nanoscale area, which is normal for electron microscopy.
The measured and modelled XRR patterns of the different metal/\gst  layer stacks are shown in Figure \ref{fig_XRR}.
In the as-deposited state, the \gst film is amorphous and there are no fringes in the measured metal-\gst samples for Ag and Al.
For comparison the corresponding theoretical model, which assume perfectly sharp interfaces, are also included in the plots.
We see that if the interfaces were ideal, as predicted by the model, then clear oscillations should be observed.
We  conclude the Al and Ag --gst{} interfaces are most likely very diffuse  as the oscillations diminish quickly.
The W/\gst XRR measurements do show oscillations but they are suppressed relative to the model predictions, which might also indicate interfacial damage between the W and \gst layers.

For Au and TiN, there are XRR fringes in the measured as-deposited state and they can be fitted with a Parratt model.
Moreover, we measured the XRR patterns of Au/\gst and TiN/\gst after annealing the structures at 100 \oc, 150 \oc, and 200 \oc~   in an argon atmosphere to prevent oxidation.
The measured  and modelled patterns are shown in Figure \ref{fig_XRR} (e) and Figure \ref{fig_XRR} (f).
We found that there are still fringes in the XRR patterns even after heating to 200 \oc, which implies that the interfaces are not damaged for Au/\gst and TiN/\gst at temperatures below 200 \oc.
However, after annealing the samples at 200~\oc, which is above the crystallisation temperature and below the temperature that Te evaporates from the film, we find that the XRR fitting improves when a 3.5~nm thick AuTe$_2$ layer is placed between the  Au and \gst layers.
For further details we encourage the interested reader to see the supporting information.

\begin{figure}[htbp]
   \centering
   \includegraphics[width=0.8\columnwidth] {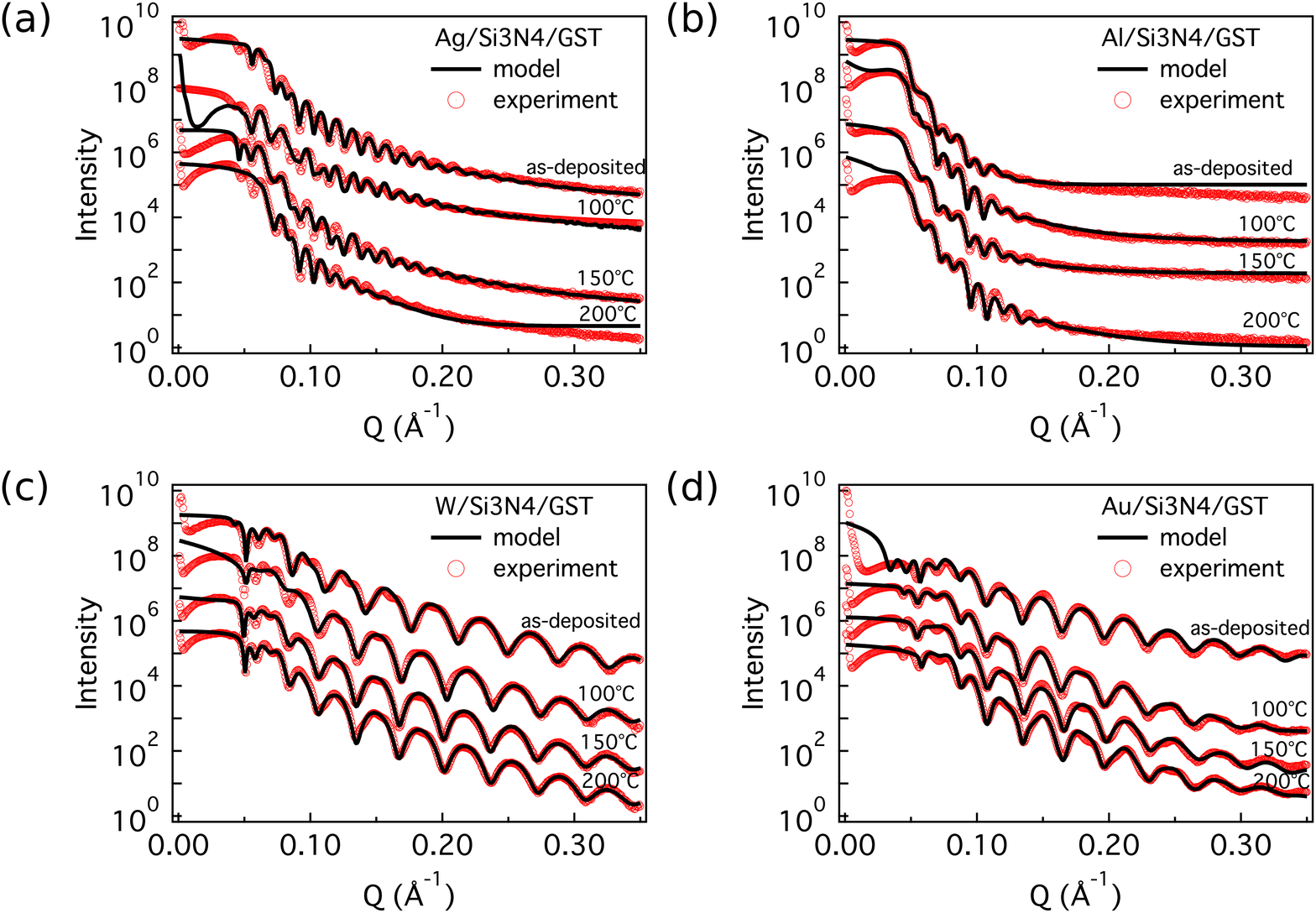} 
   \caption{Measured and modelled XRR patterns  for (a) Ag/\snns/\gstns, (b) Al/\snns/\gstns, (c) W/\snns/\gst and (d) Au/\snns/\gst in the as-deposited state and after heating to 100~\oc, 150~\oc~ and 200~\oc.}
   \label{fig_XRR_SiN}
\end{figure}

In order to prevent  inter-diffusion, a 5 nm thick \sn diffusion barrier layer was deposited between the Ag/Al and \gst{} layers.
The XRR patterns of metal-\sn-\gst samples were measured in the as-deposited state and after heating to  100 \oc, 150 \oc, and 200 \oc~ in an argon atmosphere.
The measured and modelled XRR patterns are shown in Figure \ref{fig_XRR_SiN}.
After adding the \sn barrier, we see fringes  in the measured XRR pattern.
This indicates that \sn prevented interfacial diffusion  between the metal and \gst layers.
The fringes were observed even after heating up to 200~\oc, and this suggests that the metal-\sn-\gst{} interfaces are stable.
We see that there is a difference in the XRR oscillation frequency after heating the structure above 150~\oc. 
This is due to the \gst{} amorphous-FCC phase transition, which occurs at approximately 150 \oc, and causes the \gst layer to become thinner (denser)\cite{Njoroge02}.

\subsection{Inter-diffusion and reactions between Au and \gst}
\label{sec:6}
A number of recent publications report photonic structures wherre   Au is directly interfaced with \gst{} to form a tuneable plasmonic device. 
Therefore we further investigated the inter-diffusion of Au/\gst{} using optical reflection spectroscopy, modelling the optical reflection spectrum, measuring XRR of the interface, and measuring X-Ray Diffraction (XRD) from the crystal structure.

\begin{figure}[htbp] 
  \centering
  \includegraphics[width=\columnwidth]{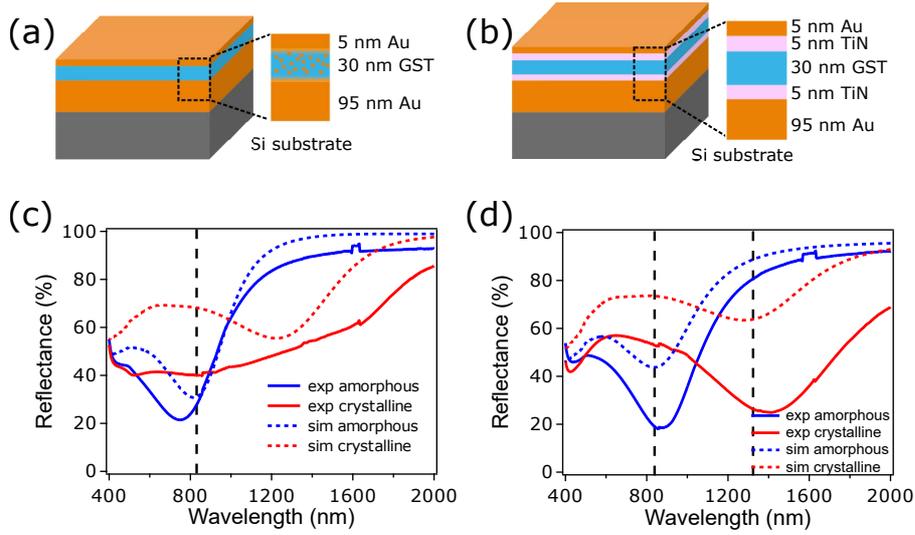} 
   \caption{(a) Schematic of Au/\gstns/Au structure without diffusion barrier, showing there is diffusion of Au atoms into \gstns. (b) Schematic of Au/TiN/\gstns/TiN/Au structure, with TiN as a diffusion barrier, depicting that TiN prevents the diffusion of Au into \gstns. The experimental and simulated reflectance spectra (c) without diffusion barrier, and (d) with TiN as a diffusion barrier.}
   \label{Reflectance}
\end{figure}

To show the effect of Au diffusion on phase change photonics devices, we fabricated a Au/\gstns/Au structure without and with a TiN diffusion barrier.
We crystallised the \gst{} by annealing the device in an argon atmosphere at 180~\oc~ for 30~minutes. 
The schematic of the structures are shown in Figure \ref{Reflectance} (a) and Figure \ref{Reflectance} (b).
The measured and simulated reflectance spectra without the diffusion barrier are shown in Figure \ref{Reflectance} (c).
In the amorphous state, we measured a strong reflective resonance at a wavelength of 744 nm, while the simulated result shows the resonance at a wavelength of 808 nm.
The difference in intensity is likely due to discrepancies in the physical constants of \sn.
Heating the Au/\gst{} structure causes the Au atoms to diffuse into \gst layer.
Indeed, there is a large discrepancy between the measured and simulated reflectance spectra after annealing the structure into the  crystalline state.
In the crystalline state, the simulated spectrum shows a resonance at a wavelength of 1209 nm, however, there is no resonance in the measured spectrum.
In contrast, when a 5 nm thick TiN diffusion barrier is used to separate the Au and \gst layers, there is good agreement between the measured and simulated spectra for both the amorphous and crystalline states of \gstns. 
This is shown in Figure \ref{Reflectance} (d).
When \gst is in the amorphous state, the measured and simulated resonances of the Au/TiN/\gstns/TiN/Au structure are at the wavelength of 864 nm and 831 nm respectively. 
After annealing the \gstns it crystallises. 
Crystallisation causes the resonances to red-shift to a wavelength of 1272~nm and 1392~nm for both the measured and simulated structures respectively.
The small differences between the simulated and measured spectra are mostly likely due to differences in the TiN dielectric function.
The small glitches at 800 nm and 1600 nm in the measured spectra shown in Figure \ref{Reflectance} (c) and Figure \ref{Reflectance} (d) occur at wavelengths where the spectrometer's detector and monochromator gratings are switched.

\begin{figure}[htbp] 
   \centering
   \includegraphics[width=\columnwidth]{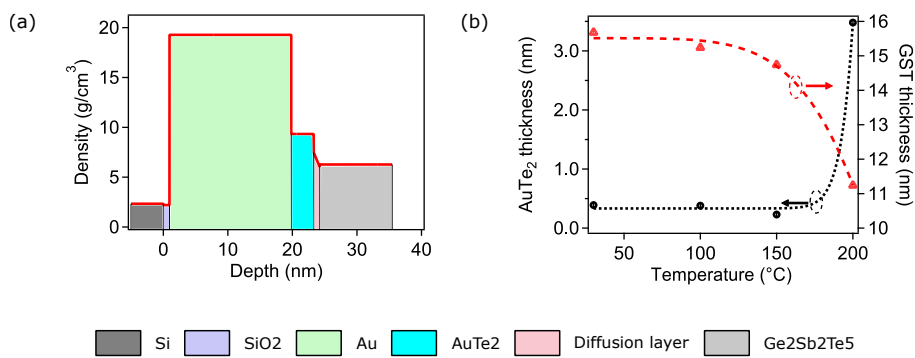} 
   \caption{(a) The modelled diffusion profile of Au/\gst annealed at 200 \oc. (b) the thickness of AuTe$_2$ layer and \gst layer as a function of the annealing temperature.}
   \label{fig_depth_profile}
\end{figure}

To unequivocally show that the change in the reflectance spectrum is due to inter-diffusion between the Au and \gstns, we used a genetic algorithm to fit the XRR pattern with a model that includes a AuTe$_2$ layer\cite{diffrac2006leptos} at the interface.
The model structure consists of a silicon substrate, a thermal silicon dioxide layer, a Au layer, a AuTe$_2$ layer, a diffusion layer, and a \gst{} layer.
The diffusion layer is modelled as a gradient layer with the density changing from the density of AuTe$_2$ to the density of \gstns.
The thickness and density of the fitted model layers after annealing at 200~\oc~ is shown in Figure \ref{fig_depth_profile} (a).
Similar plots for other annealing temperatures can be found in the supporting information.

The only way we could achieve a good fit between the measured and modelled XRR patterns was to include an interfacial Au-\gst{} layer.
The inter-diffusion of Au in to \gst increases the mass density of the interfacial layer at the interface.
The thickness of the Au/\gst{} diffusion layer, AuTe$_2$ layer, and \gst layer in the as-deposited state and after annealing at 100 \oc, 150 \oc, and 200 \oc~ are presented in the supporting information.
Figure \ref{fig_depth_profile}(b) shows the thickness of the AuTe$_2$ and \gst layers as a function of the annealing temperature.
Even after annealing the amorphous film at 100~\oc, it is clear that there is inter-diffusion between Au and \gstns.
After increasing the annealing temperature of the sample from 100\oc~  to 200\oc, the thickness of the AuTe$_2$ layer increases from 0.38 $\pm$ 0.05 nm to 3.48 $\pm$ 0.05 nm, while the thickness of the \gst decreases from 15.24 $\pm$ 0.05 nm to 11.24 $\pm$ 0.05 nm.
This indicates that Au atoms are thermally activated to diffuse further into the \gst{} layer and form the AuTe$_2$ alloy.
A logistic function has been fitted to Figure \ref{fig_depth_profile} (b) to guide the reader's eye.

\begin{figure}[htbp] 
   \centering
   \includegraphics[width=0.6\columnwidth]{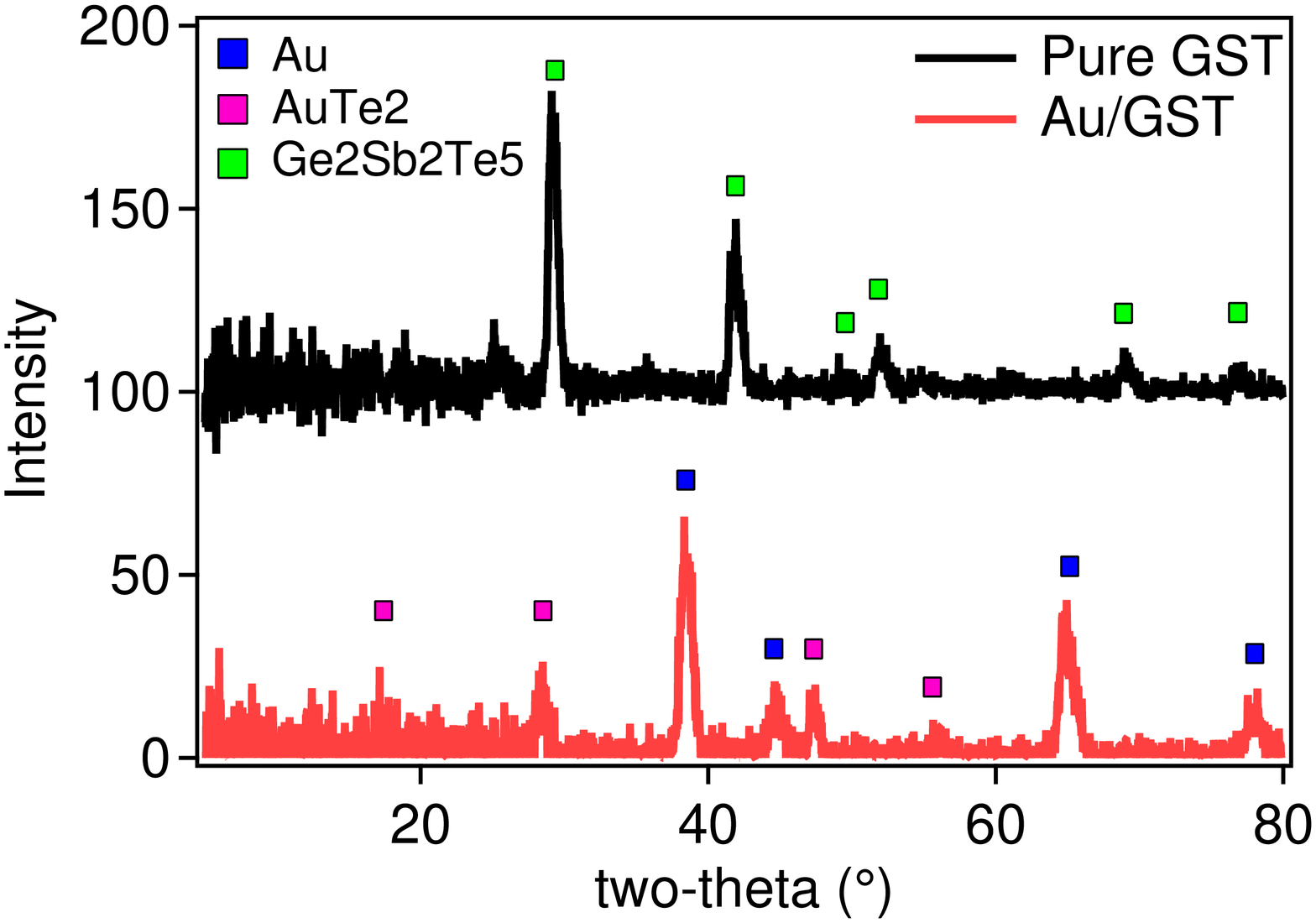} 
   \caption{XRD patterns from  Au/\gst{} and pure \gst{} after annealing at 200 \oc.}
   \label{fig_Au_XRD}
\end{figure}

The lack of resonances in the optical reflectivity spectra strongly suggests that the Au--\gst{} interface is damaged after heating. 
In contrast, the XRR pattern shows interference fringes, which suggests sharp and non-diffuse interfaces. 
These two results may suggest that the Au and \gst{} layers reacted to form a new interfacial layer with a different composition and density to that of \gst{} and Au. 
Therefore, we measured the X-ray diffraction (XRD) pattern from the Au/\gst{} structure after annealing at 200 \oc to identify crystallographic  changes to the \gst{} structure, see Figure \ref{fig_Au_XRD}.

It is clear from figure \ref{fig_Au_XRD} that the Au layer produces additional peaks in the \gst{} spectra. 
By comparing the XRD patterns of Au \cite{ellner1991partial}, AuTe$_2$ \cite{reithmayer1993high}, and face centred cubic \gst in the crystalline state \cite{jeong2000crystal}, we see that the diffraction pattern has contributions from Au, AuTe$_2$, and \gstns.
This shows that \gst and Au reacted  to form Au$Te_2$.
Since chemical reactions are activated processes, the likelihood of AuTe$_2$ forming depends on the Au concentration. 
Others have shown that for films thicker than 100~nm, Au does not diffuse into GeTe. 
In contrast, our XRD and XRR measurements suggest that when the Au--\gst{} (20~nm and 20~nm) structure is heated  to just 200~\oc,   a 3.5~nm thick AuTe$_2$ layer forms at the interface.
This interface corresponds to 17.5\% of the 20~nm film thickness, which may explain why we are able to detect this problem. 
We suspect that if the 100~nm thick GeTe-Au structures had been heated for a longer period time at higher temperatures, then this Au alloying problem would have been observed.
Considering that most optical structures use very thin layers of \gst{}\cite{Wuttig2017NP}, it is important to consider interfacial diffusion and reactions.

Due to the formation of AuTe$_2$, the Ge-Sb-Te composition moves to a Ge rich section of the compositional space, where the crystallisation temperature is greater than that of \gstns\cite{cheng2011high}.
 In addition, the TiN and \sn diffusion barriers between the metal and \gst can also induce stress on the \gst film,  which can influence its crystallisation temperature\cite{simpson2009toward}.
Since the crystallisation temperature depends on the interfacial reactions, the PCM composition, and the mechanical properties of the materials used in the layered structure, the affect of different metallic caps and diffusion barriers on the crystallisation kinetics needs to be studied carefully.

We also measured the XRD pattern of TiN/\gst after annealing at 200 \oc~ and this can be found in the supporting information, see Figure S2.
The XRD pattern of TiN/\gst{} is similar to that of \gst{} in the crystalline state, which means TiN is stable and does not diffuse into \gst at a temperature of 200~\oc.
Indeed, the thermal stability of \gst{} in contact with Ti and TiN has been previously investigated \cite{venugopal2009thermal}. 
TiN layers do not react with the \gst{} film, even after annealing at 450 \oc. 
However, TiTe$_2$ is formed when Ti interfaces with \gst{} after annealing at 300 \oc. 
Note,  Ti is sometimes used to increase the adhesion of TiN in phase change memory devices \cite{venugopal2012compatibility}. 
The interested reader is referred to a series of studies that discuss the use of Ti-based materials diffusion barriers in phase change memory devices\cite{alberici2004ti,loubriat2011gete,park2016phase}.

We have optically and structurally that  Au--\gst{} interfaces are unstable and that adding an  thin interfacial TiN layer between the Au and \gstns, prevents interfacial reaction problems.
Even in the published literature there are discrepancies between the simulated and measured properties of tuneable photonic devices using metals and phase change materials \cite{jafari2017ultra,jafari2017ultra2,michel2013using}.
We believe that inter-diffusion of metals into phase change materials may be the reason for these discrepancies.
We also expect that inter-diffusion may effect the performance of other chalcogenide photonics systems, such as quantum dots placed on metallic electrodes \cite{yarema2011infrared}.
However, we have also shown a simple solution.
We found that measurements of  optical and x-ray reflectivity resonances remain even after thermally switching the \gst{} layer when  a TiN diffusion barrier  is added between the \gst{} and the Au.
We also showed that the TiN layer also prevents interfacial chemical reactions between Au and \gst{} layers.  
Indeed, TiN is commonly interfaced directly with \gst{} in phase change random access memory devices, and they can be cycled millions of times \cite{burr2010phase}.
Since TiN is also a plasmonic material, and is electrically conductive, we believe it is well suited to electrically reprogrammable phase change plasmonic devices.
With these points in mind, we strongly recommend that those designing new tuneable plasmonic devices, which are based on Telluride phase change materials, to account for diffusion barrier layers in their designs. 
Or better still, replace the plasmonic metal layer with TiN.

\section{Conclusion}
\label{con}
In conclusion, there is substantial interfacial damage when Au, Ag,  and Au  plasmonic metals are directly interfaced with phase change chalcogenide layers. 
Diffusion occurs at room temperature and is accelerated when the phase change material is heated during the switching process.
However,  diffusion barriers, such as \sn and TiN can prevent the inter-diffusion. 
Atomic diffusion  and interfacial chemical reactions change the composition of the Ge-Sb-Te layer, which influences its crystallisation temperature and concomitant  optical response.
We found that the \sn diffusion barrier prevents inter-diffusion to at least 200 \oc~ and we expect the barrier to be stable at higher temperatures.

Our results show that Au inter-diffuses with \gst{} and therefore should be avoided in tuneable plasmonic device designs. 
However, TiN, which is also a plasmonic material with a similar dielectric function to Au does not diffuse into \gst.
Our experiments show that TiN can directly contact \gst{} without diffusion even after heating to temperatures greater than 200 \oc.
Considering that TiN is also a proven heating electrode in phase change memory devices, we expect the interfaces to be stable up to the \gst{} melting temperature of 650 \oc.  
These two factors strongly suggest that TiN should be used in practical electrically tuned phase change plasmonic devices. 
In short, we strongly recommend that those designing phase change photonics devices  consider the diffusion between phase change materials and metal layers, and where possible replace Au with TiN.

\begin{acknowledgements}
This work was performed under the auspices of the SUTD-MIT international design center (IDC) with project funding from A-Star (project number 1420200046), the Singapore Ministry of Education (MoE) Tier 2 (project number MOE2017-T2-1). 
The work was initiated by a Samsung GRO project. L Lu, W Dong, and J Behera are grateful for their MoE funded SUTD PhD scholarships.
\end{acknowledgements}

\section*{Conflicts of interest}
The authors declare no conflict of interest.

\section*{Supplementary material}
\begin{itemize}
  \item Diffusion profile of Au/\gst in the as-deposited state and after annealing at different temperatures
  \item XRD pattern of TiN/\gst after annealing at 200 \oc
\end{itemize}

\bibliographystyle{unsrt}

\end{document}


\title{Supporting information}
\subtitle{Inter-diffusion of Plasmonic Metals and Phase Change Materials}

\author{Li Lu \and Weiling Dong \and  Jitendra K. Behera \and Li Tian Chew \and Robert E. Simpson}

\institute{Li Lu \at
              Singapore University of Technology and Design, 8 Somapah Road, Singapore, 487372 \\
              \email{li\_lu@mymail.sutd.edu.sg} 
           \and
           Weiling Dong \at
              Singapore University of Technology and Design, 8 Somapah Road, Singapore, 487372 \\
              \email{weiling\_dong@mymail.sutd.edu.sg} 
           \and
           Jitendra K. Behera \at
              Singapore University of Technology and Design, 8 Somapah Road, Singapore, 487372 \\
              \email{jitendra\_behera@mymail.sutd.edu.sg} 
           \and
	    Li Tian Chew \at
              Singapore University of Technology and Design, 8 Somapah Road, Singapore, 487372 \\
              \email{litian\_chew@sutd.edu.sg}
          \and
          Robert E. Simpson \at
              Singapore University of Technology and Design, 8 Somapah Road, Singapore, 487372 \\
              Tel.: +65 64994519 \\
              \email{robert\_simpson@sutd.edu.sg}
}
\date{Received: date / Accepted: date}
\maketitle

\newpage
\section{Diffusion profile of Au/\gstns}

\begin{figure}[htbp] 
   \centering
   \includegraphics[width=0.8\columnwidth]{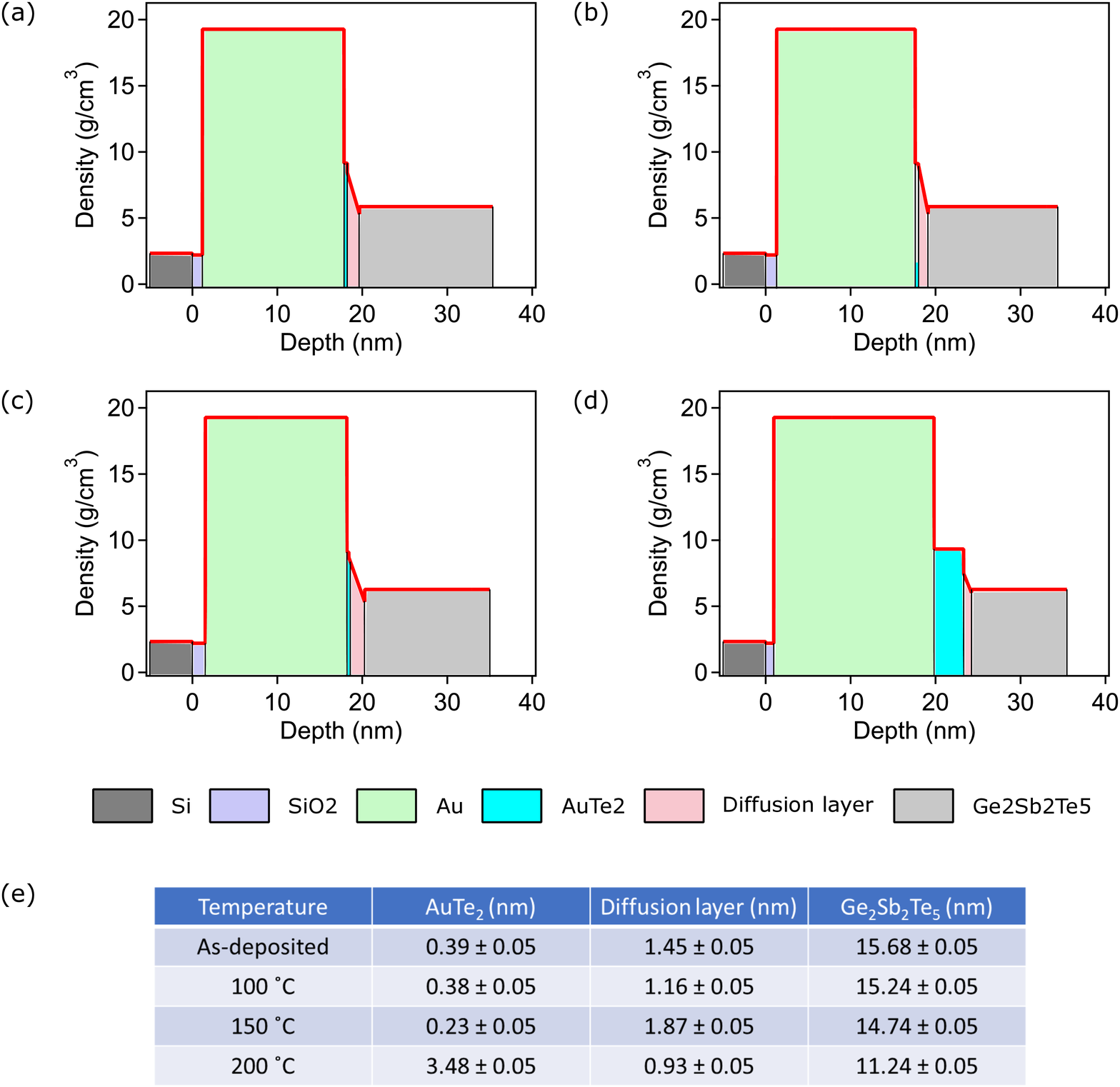} 
   \caption{The modeled diffusion profile of Au/\gst (a) in the as-deposited state, (b) annealed at 100 \oc, (c) annealed at 150 \oc, (d) annealed at 200 \oc. (e) The thickness of the diffusion layer at different temperatures.}
   \label{fig_depth_profile}
\end{figure}

The modelled diffusion profiles of Au/\gst in the as-deposited state and after annealing at 100 \oc, 150 \oc, and 200 \oc~ are shown in Figure \ref{fig_depth_profile} (a)--(d).
The thickness and roughness of the Au/\gst diffusion layer in the as-deposited state and after annealing at 100 \oc, 150 \oc, and 200 \oc~ are shown in Figure \ref{fig_depth_profile} (e).
The model structure consists of  a silicon substrate, a thermal silicon dioxide layer, a Au layer, a AuTe$_2$ layer, a diffusion layer, and a \gst layer.
The silicon dioxide layer is due to the thermal oxidation of the silicon substrate.
The diffusion layer is modelled as a gradient layer with the density changing from the density of AuTe$_2$ to the density of \gstns.

\section{XRD pattern of TiN/\gstns}

\begin{figure}[htbp] 
   \centering
   \includegraphics[width=0.6\columnwidth]{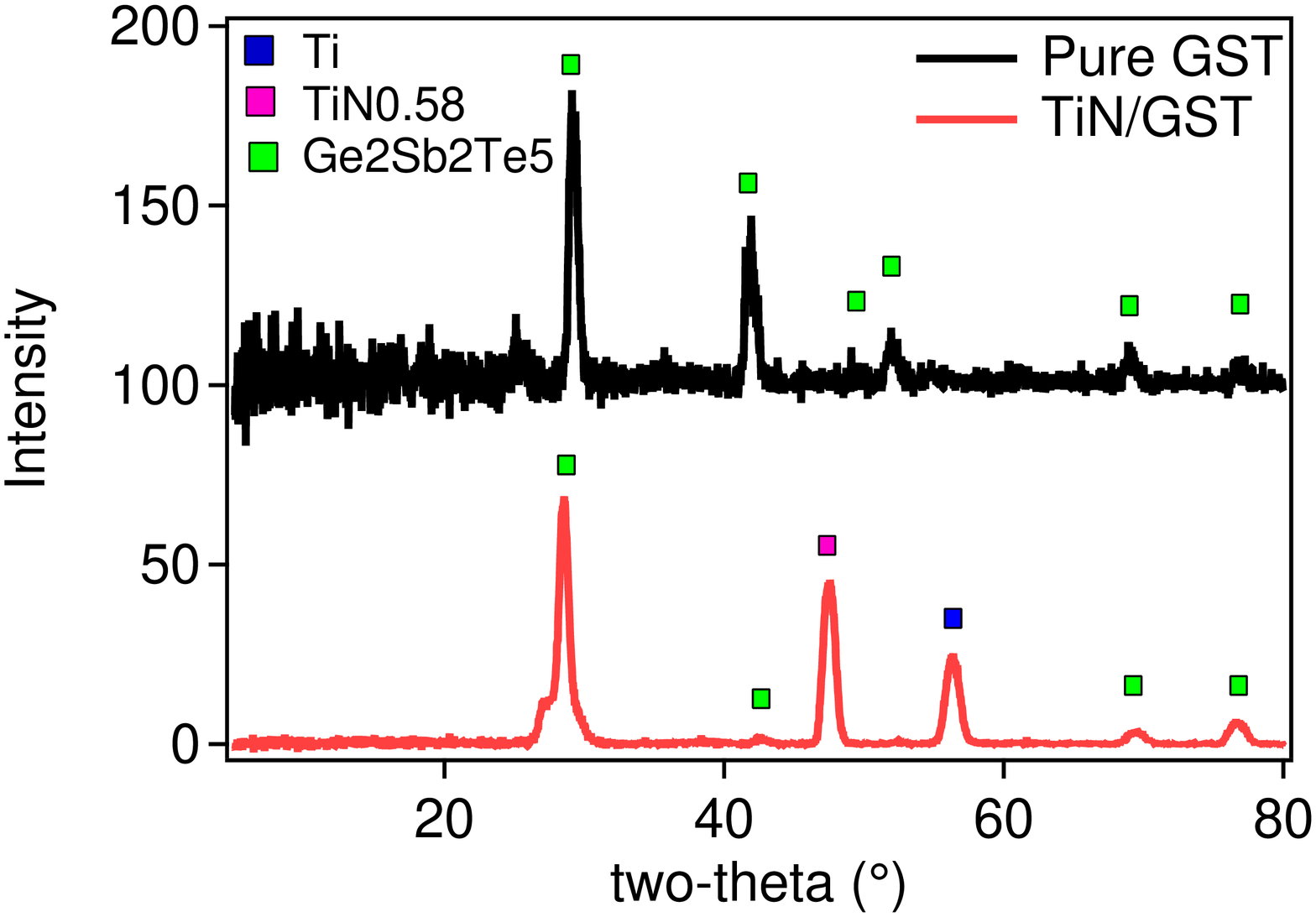} 
   \caption{XRD results of TiN/\gst and \gst after annealing at 200 \oc.}
   \label{fig_TiN_XRD}
\end{figure}

The XRD patterns from TiN/\gst and \gst after annealing at 200 \oc~ are shown in Figure \ref{fig_TiN_XRD}.
The XRD peaks of TiN/\gst are in accordance with  face centered cubic \gst in the crystalline state.
There are also Ti and TiN$_{0.58}$ peaks, which are due to the non-stoichiometric phase of titanium nitride (TiN$_x$), which is normally used for plasmonic applications.
In our case, the TiN$_x$ layer is Ti rich but  it is still stable and does not diffuse into \gst even at a temperature of 200\oc.